# Mid-infrared frequency comb with 25 pJ threshold via CW-seeded optical parametric generation in nonlinear waveguide

MIKHAIL ROIZ[1,*], JUI-YU LAI[2], JUHO KARHU[3] AND MARKKU VAINIO[1,4,5]

[1]*Department of Chemistry, University of Helsinki, FI-00560, Helsinki, Finland*
[2]*HC Photonics Corp. Hsinchu Science Park, Hsinchu 30078, Taiwan*
[3]*Metrology Research Institute, Aalto University, Espoo, FI-00076, Finland*
[4]*Photonics Laboratory, Physics Unit, Tampere University, Tampere, FI-33101, Finland*
[5]*e-mail: markku.vainio@helsinki.fi*
*Corresponding author: mikhail.roiz@helsinki.fi*

**We demonstrate efficient generation of mid-infrared frequency combs based on continuous-wave-seeded femtosecond optical parametric generation in nonlinear waveguides. Conversion of the near-infrared pump to signal and idler light takes place with very high efficiency (74 %) and the threshold (25 pJ for 100 fs pulses) is over 300 times lower than in bulk analogs. Relative intensity noise of the mid-infrared comb is exceptionally low, below $5 \times 10^{-5}$ (integrated from 10 Hz to 2 MHz). Furthermore, the mid-infrared bandwidth can be increased by driving the process with broadband pump obtained via supercontinuum generation.**

Combining high coherence and broad optical bandwidth with the possibility of linking the infrared molecular fingerprint region to radio-frequency standards, Mid-Infrared (MIR) Optical Frequency Comb (OFC) technology is a superior tool for high-resolution spectroscopy [1, 2]. Despite the fact that in the past two decades great progress has been made in the development of direct MIR OFC generation based on quantum cascade lasers [3] and mode-locked lasers (MLL) [4], parametric frequency conversion techniques still provide OFCs with larger instantaneous optical bandwidth and more versatile wavelength coverage. Along with the well-known and widely used Difference Frequency Generation (DFG) [5-7] and Optical Parametric Oscillation (OPO) [8-10], continuous-wave (CW) seeded femtosecond Optical Parametric Generation (OPG) is emerging as a promising alternative for *fully stabilized* MIR comb generation with significant advantages [11].

In addition to its robustness and simplicity, CW-seeded OPG features high conversion efficiency and excellent pulse-to-pulse coherence [11-13]. Recently, we have demonstrated that single-pass CW-seeded OPG in MgO-doped periodically poled lithium niobate (MgO:PPLN) pumped near 1 µm wavelength allows for the generation of fully stabilized MIR combs with the possibility to dynamically control the Carrier-Envelope Offset (CEO) frequency [11]. The CEO control in this case is inherently independent of the repetition rate, making the method highly versatile for applications in spectroscopy and frequency metrology, for instance in high-resolution interleaved dual-comb spectroscopy [14]. Moreover, the MIR comb's CEO determination is not required since its value is always known [11]. The seeded OPG also has low Relative Intensity Noise (RIN) as long as the seeding is done using a CW laser [11, 15]. This helps to remove additional noise due to the pulse-to-pulse fluctuations that may arise in other systems such as femtosecond DFG [16].

The implementation of CW-seeded OPG in a bulk crystal requires ultrashort pump pulses with energies at the nJ level, which needs amplification. Additionally, the instantaneous optical bandwidth is rather limited due to the requirement for a long nonlinear crystal [13, 17]. In this Letter, we report a solution that significantly improves the OPG comb performance in terms of power efficiency, threshold, RIN and instantaneous optical bandwidth. First, we take advantage of state-of-the-art nonlinear waveguide technology and achieve over 300-fold reduction in the CW-seeded OPG threshold compared to bulk analogs [11, 13, 17] and over 10-fold threshold reduction compared to previous waveguide solutions without CW seeding [18]. The reached pJ level threshold corresponds to just a few milliwatts of average pump power, which is readily achievable with standard mode-locked lasers. Furthermore, the obtained total energy conversion efficiency of up to 74 % is exceptionally high. This opens up opportunities for miniaturization of the system and allows lasers with low energy pulses to efficiently drive the OPG, which is especially relevant to high repetition rate light sources [19]. Second, despite using a long waveguide, we demonstrate that the CW-seeded OPG driven by a pump broadened via supercontinuum (SC) generation can significantly increase the instantaneous optical bandwidth of the MIR comb with only a small change in the OPG threshold. Overall, we anticipate that the approach presented in this work is an important step towards compact, power-efficient and fully stabilized mid-infrared frequency comb generators.

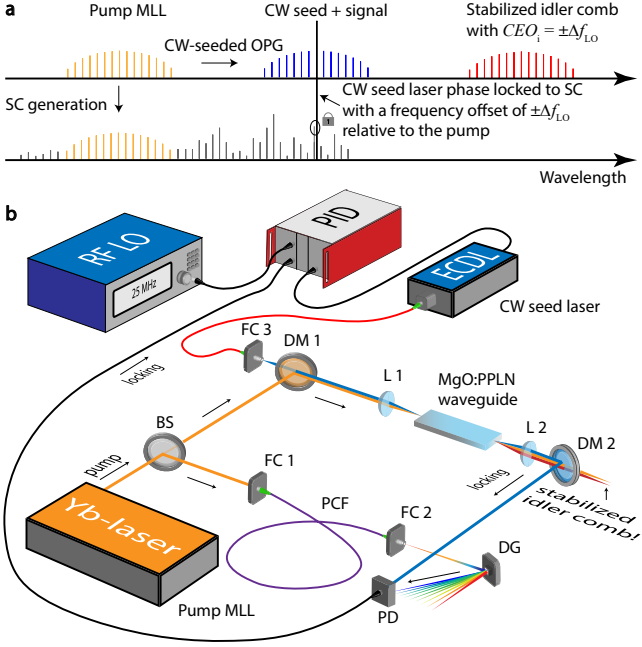

Fig. 1. (a) Schematic of the CW-seeded OPG. (b) Experimental setup. BS: beam splitter, FC: fiber collimator, DG: diffraction grating, DM: dichroic mirror, L: lens, PD: photodiode, PID: Proportional–Integral–Derivative controller, ECDL: External Cavity Diode Laser, PCF: Photonic Crystal Fiber; Components with secondary importance are omitted for simplicity.

In general, a frequency comb consists of multiple equidistantly spaced laser modes, whose optical frequencies $v_n$ can be described using two radio frequencies [20]:

$$v_n = CEO + n \cdot f_r,$$

where $f_r$ is the repetition rate, n is an integer mode number and CEO is the carrier-envelope offset frequency. In the OPG process each pump pulse generates signal and idler pulses, and the repetition rates of all three pulse trains are equal. On the other hand, the CEO of the signal and idler pulses vary randomly from pulse to pulse, since OPG is based on the amplification of quantum noise. The CEO frequencies in the non-degenerate OPG process can be described by the following equation:

$$CEO_p = CEO_s + CEO_i, \qquad (1)$$

where $CEO_p$, $CEO_s$ and $CEO_i$ are the offset frequencies of the pump, signal and idler combs, respectively. As evident from equation (1), $CEO_s$ and $CEO_i$ may take different values to fulfill the equation, which leads to pulse-to-pulse incoherence. Recently, using a bulk MgO:PPLN crystal, we have shown that the idler CEO can be stabilized by seeding the OPG with a narrow-linewidth CW laser referenced to the pump MLL [11]. Here, we have improved the experimental setup (see Supplement 1 for details) and demonstrated that this approach can be made compatible with low-power pump lasers by using nonlinear waveguides. The corresponding schematic of the CW-seeded OPG concept for fully stabilized MIR comb generation in waveguides as well as the experimental setup used in this work can be seen in Figs. 1a and 1b, respectively. The light from a pump MLL is split into two arms. One of them is used to drive the OPG, and the other is used to produce a SC serving as a reference for the CW seed laser. The generated SC should reach the CW seed laser wavelength so that a beat note between the CW seed laser and the SC is produced, which is then used for phase-locking. When the phase-locking and seeding are performed, the generated signal comb shares the same CEO as the pump MLL with an additional frequency offset $\pm \Delta f_{LO}$. In this case, the CEO equation can be written as follows:

$$CEO_p = CEO_s + CEO_i = (CEO_p \pm \Delta f_{LO}) + CEO_i \qquad (2)$$

meaning that $CEO_i$ is equal to either $-\Delta f_{LO}$ or $\Delta f_{LO}$. The frequency offset $\Delta f_{LO}$ is defined by the radio frequency (RF) local oscillator (LO) involved in the phase-locking procedure, hence it can be used to control the idler CEO independent of the repetition rate. It is worth noting that in this configuration the $CEO_p$ *does not need active stabilization*, since the phase-locked seed laser will automatically readjust the $CEO_s$ and ensure the absolute stability of $CEO_i$ in case $CEO_p$ drifts or fluctuates.

Our experimental setup includes a 1042-nm Yb-doped fiber MLL (MenloSystems GmbH, Orange comb FC1000-250) that generates 100 fs Gaussian pulses at 250 MHz repetition rate. The $CEO_p$ is *free running*. The comb tooth linewidth (FWHM) of <200 kHz, measured at 100 ms timescale. The repetition rate is locked to an RF source; however, the use of an optical reference (stable CW laser) can reduce the linewidth of the comb teeth [14]. In order to generate a suitable SC, we used 80 cm long photonic crystal fiber (NKT Photonics, NL-PM-750) that requires about 70 – 100 mW of pump power inside the fiber [11]. The CW seed laser is a commercial external cavity diode laser (Toptica Photonics, CTL 1550) that is phase-locked to the SC (see Fig. 1b). The waveguide was fabricated by HC Photonics. It is a MgO:PPLN ridge waveguide made by bonding, etching and thinning [21]. It is able to sustain high power for pulse applications as compared to traditional proton exchanged waveguides. The waveguide has a 12 mm long quasi-phase-matching uniform grating with a period of 21.71 µm and a mode field diameter of 5.6 x 4.3 µm. The input polarizations to the waveguide as well as to the photonic crystal fiber were controlled by half-wave plates.

In Fig. 2 one can see the optical spectra generated in the CW-seeded OPG setup. Note that there is an additional sum-frequency generation process [18]. One could argue that the CW-seeded OPG presented here is just a version of an Optical Parametric Amplifier (OPA). However, OPAs do not usually have threshold. In contrast, our system exhibits a clear threshold behavior with and without CW seeding. Fig. 3a shows how the input pump power converts to signal and idler power without seeding and with 1.8 mW CW seed power inside the waveguide. The lowest threshold of just 25 pJ (6.3 mW average pump power) is reached with CW seeding, corresponding to over two orders of magnitude improvement to the previously reported bulk systems [11, 13, 17] and an order of magnitude improvement compared to the earlier OPG experiments in quasi-phase-matched lithium niobate waveguides [18]. Recently there have been reports on OPG and OPA in new type of nanophotonic lithium niobate waveguides [22, 23]. The reports demonstrate threshold of just 4 pJ, which is remarkably low for the OPG process in general. However, the reported total conversion efficiency is limited to about 10 %, while we reach 74 % (23 % pump-to-idler conversion efficiency), representing the state-of-the-art [18, 22].

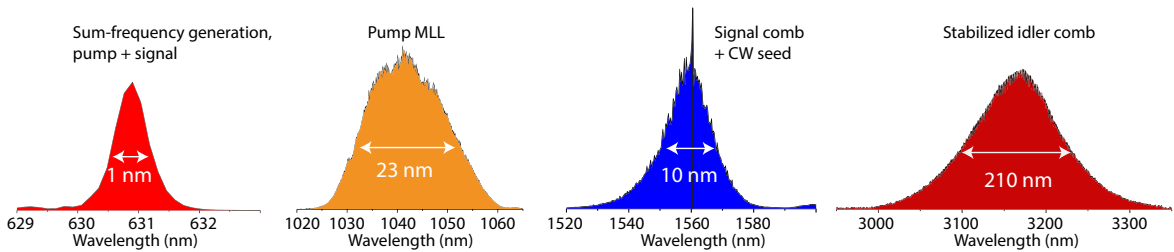

Fig. 2. CW-seeded OPG experimental spectra shown in the linear scale with arbitrary units.

Note that our conversion efficiency was calculated from the signal and idler output powers that were measured after a collimating lens and dichroic optics, meaning that the internal photon-conversion efficiency is even higher. It is also worth emphasizing that the coupling efficiency of the pump light to the waveguide was over 45% using free-space coupling. The OPG process saturates at pump pulse energies above 110 pJ or 27.5 mW of average pump power, which sets a limit to the maximum idler average power of 5.8 mW. This power is enough for most applications, such as high-resolution dual-comb spectroscopy and comb-assisted spectroscopy. These applications do not require high powers, but instead high coherence and stability of the MIR comb [8, 14]. In Fig. 3b it is demonstrated how CW seed power changes the total converted signal and idler pulse energies when the pump energy is fixed at 36 pJ (9 mW average power), which is lower than the pulse energy required to start the OPG without seeding. It is clear that for the seed powers larger than 1.8 mW the output power does not increase any further. For that reason, we kept this seed power for the rest of the measurements.

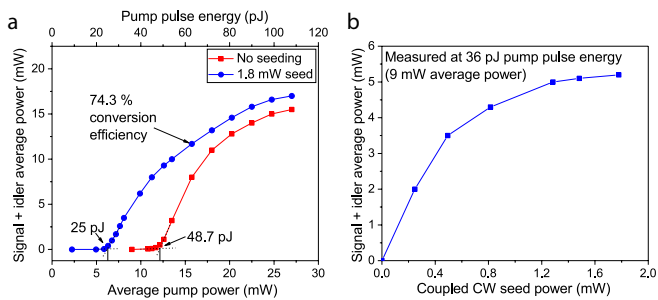

Fig. 3. (a) Total converted signal + idler power vs. pump power without seeding (red squares) and with 1.8 mW seed power (blue circles). (b) CW seed power vs. signal and idler average power.

An important additional benefit of the waveguide-based implementation is the possibility to pump the OPG directly with a low-noise MLL. This way we avoid excess intensity noise generated in an optical amplifier, and as a result we have reached exceptionally low RIN. The RIN was measured with and without CW seeding in the saturated and non-saturated regimes (see Fig. 4a) using method described in [24]. CW seeding decreases the noise drastically and the root-mean-square RIN of the mid-infrared comb becomes smaller than $5\times10^{-5}$ when integrated from 10 Hz to 2 MHz, where the noise is mostly below the detection limit. This noise level is substantially smaller than what has been reported for CW-seeded OPG [11, 15] or, for example, state-of-the-art DFG combs [16, 25, 26].

Next, we proved the idler CEO stability via a comparison to another, independently stabilized OFC (out-of-loop measurement). We combined the idler second harmonic with an independent reference comb with the same repetition rate (fully stabilized Er-doped fiber comb, MenloSystems GmbH, FC1500-250-WG) to produce a relative offset beat note. After that, we performed frequency counting of the obtained beat note; the results can be seen in Fig. 4b proving the high accuracy and stability of $CEO_i$. Note that the high measurement noise (the standard deviation in Fig. 4b) is present simply due to the relative instability (timing jitter) between our Er-doped (reference) and Yb-doped (pump) MLLs. The repetition rates of the lasers are locked to two different RF signal generators that are both referenced to the same GPS-disciplined crystal oscillator with the specified relative instability of $5\times10^{-12}$ in 1 s. In order to get rid of this limitation, we measured $CEO_i$ against another MIR comb, which was generated via CW-seeded OPG in bulk crystal using the same pump MLL as for the waveguide OPG process; see Supplement 1 for more information. The seed lasers of the two OPG setups were locked to the pump via the same supercontinuum. The relative offset beat note between the two mid-infrared idler combs can be seen in Fig. 4c with sub-Hz linewidth, limited by the RF spectrum analyzer. By performing frequency counting of this beat note (Fig. 4d), we can see excellent relative stability of the two combs. The noise averages out as white phase noise, confirming the absence of any drifts.

Common methods for obtaining broad MIR spectra from nonlinear comb generators (OPO, DFG, OPG) include the use of short [10] or chirped QPM structures [5, 27, 28]. These both widen the phase-matching bandwidth but at the cost of increased threshold. Here, we show that the MIR bandwidth can be substantially increased without significant gain reduction by simply increasing the pump optical bandwidth. We used just 3.7 cm of photonic crystal fiber (NKT Photonics, NL-PM-750) to broaden the pump spectrum from 23 nm to about 200 nm (see Supplement 1 for more details) with minimum distortions of the pulse in time domain. We used the broadened pump directly for the CW-seeded OPG with no additional pulse compression stage. For comparison, in Fig. 5a one can see the idler spectrum without pump broadening. The full detectable optical bandwidth in this case is about 300 nm. If the pump is broadened, the resulting MIR bandwidth can be at least doubled, and the spectral shape can be varied by changing the photonic crystal fiber input polarization (see Figs. 5b-d). Importantly, the threshold increases by only a small amount of 15-20 % depending on the input polarization state to the photonic crystal fiber. In view of further miniaturization of the experimental implementation, the photonic crystal fiber can be replaced, e. g., by MgO:PPLN [29, 30], $Si_3N_4$ [25, 31] or silicon waveguides [26].

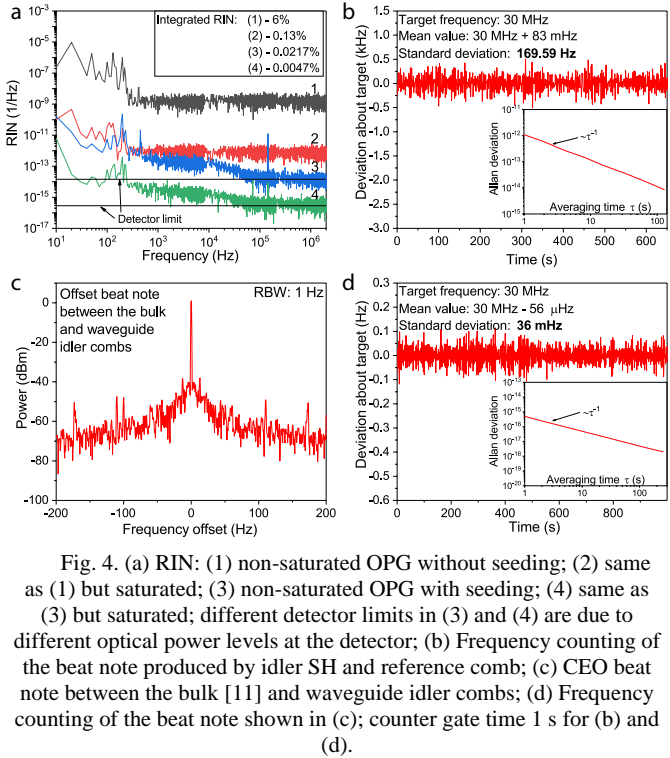

Fig. 4. (a) RIN: (1) non-saturated OPG without seeding; (2) same as (1) but saturated; (3) non-saturated OPG with seeding; (4) same as (3) but saturated; different detector limits in (3) and (4) are due to different optical power levels at the detector; (b) Frequency counting of the beat note produced by idler SH and reference comb; (c) CEO beat note between the bulk [11] and waveguide idler combs; (d) Frequency counting of the beat note shown in (c); counter gate time 1 s for (b) and (d).

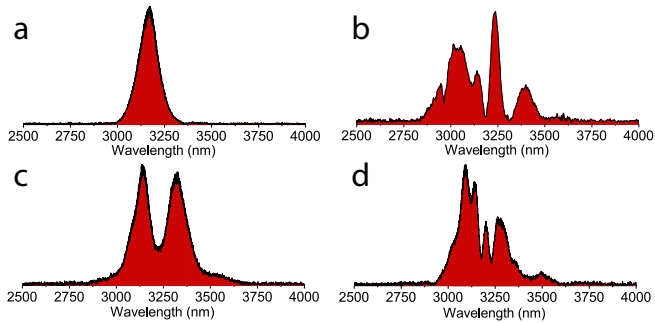

Fig. 5. (a) idler spectrum without the pump broadening. (b-d) The idler spectra with pump broadened in 3.7 cm photonic crystal fiber;

In conclusion, we have demonstrated a new approach for fully stabilized MIR OFC generation using CW-seeded OPG in nonlinear waveguides. We have achieved OPG with extremely high conversion efficiency of 74 % in a single-pass configuration. Both the threshold (25 pJ) and the MIR comb RIN ($5 \times 10^{-5}$) are exceptionally low. Moreover, the generated MIR comb features dynamic CEO control independent of the repetition rate and it does not need an additional measurement setup to determine its absolute value. This platform paves the way towards efficient high repetition rate MIR OFC generation, which is especially interesting for fast spectroscopic measurements, for instance, in reaction kinetics [2].

**Funding.** The work was funded by the Academy of Finland (Project numbers 326444 and 314364).

**Acknowledgment.** Authors wish to thank Dr. Christopher Phillips from ETH Zurich for valuable discussions on the manuscript.

**Disclosures.** The authors declare no conflicts of interest.

**Data availability.** Data underlying the results presented in this paper are not publicly available at this time but may be obtained from the authors upon reasonable request.

**Supplemental document.** See Supplement 1 for supporting content.

**References**

1. A. Schliesser, N. Picqué, and T. W. Hänsch, Nat. Photon. **6,** 440-449 (2012).
2. M. Vainio and L. Halonen, Phys. Chem. Chem. Phys. **18,** 4266-4294 (2016).
3. L. Consolino, M. Nafa, F. Cappelli, K. Garrasi, F. P. Mezzapesa, L. Li, A. G. Davies, E. H. Linfield, M. S. Vitiello, P. De Natale, and S. Bartalini, Nat. Commun. **10,** 2938 (2019).
4. S. Vasilyev, I. Moskalev, V. Smolski, J. Peppers, M. Mirov, V. Fedorov, D. Martyshkin, S. Mirov, and V. Gapontsev, Optica **6,** 126-127 (2019).
5. L. Zhou, Y. Liu, H. Lou, Y. Di, G. Xie, Z. Zhu, Z. Deng, D. Luo, C. Gu, H. Chen, and W. Li, Opt. Lett. **45,** 6458-6461 (2020).
6. T. W. Neely, T. A. Johnson, and S. A. Diddams, Opt. Lett. **36,** 4020-4022 (2011).
7. S. Vasilyev, I. S. Moskalev, V. O. Smolski, J. M. Peppers, M. Mirov, A. V. Muraviev, K. Zawilski, P. G. Schunemann, S. B. Mirov, K. L. Vodopyanov, and V. P. Gapontsev, Optica **6,** 111-114 (2019).
8. A. V. Muraviev, V. O. Smolski, Z. E. Loparo, and K. L. Vodopyanov, Nat. Photon. **12,** 209-214 (2018).
9. M. Vainio and J. Karhu, Opt. Express **25,** 4190-4200 (2017).
10. N. Leindecker, A. Marandi, R. L. Byer, K. L. Vodopyanov, J. Jiang, I. Hartl, M. Fermann, and P. G. Schunemann, Opt. Express **20,** 7046-7053 (2012).
11. M. Roiz, K. Kumar, J. Karhu, and M. Vainio, APL Photonics **6,** 026103 (2021).
12. V. Petrov, F. Noack, and R. Stolzenberger, Appl. Opt. **36,** 1164-1172 (1997).
13. A. Aadhi and G. K. Samanta, Opt. Lett. **42,** 2886-2889 (2017).
14. A. V. Muraviev, D. Konnov, and K. L. Vodopyanov, Sci. Rep **10,** 18700 (2020).
15. W. Chen, J. Fan, A. Ge, H. Song, Y. Song, B. Liu, L. Chai, C. Wang, and M. Hu, Opt. Express **25,** 31263-31272 (2017).
16. V. Silva de Oliveira, A. Ruehl, P. Masłowski, and I. Hartl, Opt. Lett. **45,** 1914-1917 (2020).
17. H. Linnenbank and S. Linden, Opt. Express **22,** 18072-18077 (2014).
18. X. Xie, A. M. Schober, C. Langrock, R. V. Roussev, J. R. Kurz, and M. M. Fejer, J Opt Soc Am B **21,** 1397-1402 (2004).
19. A. S. Kowligy, D. R. Carlson, D. D. Hickstein, H. Timmers, A. J. Lind, P. G. Schunemann, S. B. Papp, and S. A. Diddams, Opt. Lett. **45,** 3677-3680 (2020).
20. T. Fortier and E. Baumann, Commun. Phys. **2,** 153 (2019).
21. Cheng-Wei Hsu, Jui-Yu Lai, Chen-Shao Hsu, Yu-Tai Huang, K. Wu, and Ming-Hsien Chou, Proc. SPIE **10902,** (2019).
22. M. Jankowski, N. Jornod, C. Langrock, B. Desiatov, A. Marandi, M. Lončar, and M. M. Fejer, arXiv, 2104. 07928 (2021).
23. L. Ledezma, R. Sekine, Q. Guo, R. Nehra, S. Jahani, and A. Marandi, arXiv, 2104. 08262 (2021).
24. R. P. Scott, C. Langrock, and B. H. Kolner, IEEE J. Sel. Top. Quantum Electron. **7,** 641-655 (2001).
25. A. Mayer, C. Phillips, C. Langrock, A. Klenner, A. Johnson, K. Luke, Y. Okawachi, M. Lipson, A. Gaeta, M. Fejer, and U. Keller, Phys. Rev. Applied **6,** 054009 (2016).
26. N. Nader, D. L. Maser, F. C. Cruz, A. Kowligy, H. Timmers, J. Chiles, C. Fredrick, D. A. Westly, S. W. Nam, R. P. Mirin, J. M. Shainline, and S. Diddams, APL Photonics **3,** 036102 (2018).
27. M. Charbonneau-Lefort, B. Afeyan, and M. M. Fejer, J Opt Soc Am B **25,** 1402-1413 (2008).
28. C. R. Phillips, B. W. Mayer, L. Gallmann, M. M. Fejer, and U. Keller, Opt. Express **22,** 9627-9658 (2014).
29. C. R. Phillips, C. Langrock, J. S. Pelc, M. M. Fejer, I. Hartl, and M. E. Fermann, Opt. Express **19,** 18754-18773 (2011).
30. M. Jankowski, C. Langrock, B. Desiatov, A. Marandi, C. Wang, M. Zhang, C. R. Phillips, M. Lončar, and M. M. Fejer, Optica **7,** 40-46 (2020).
31. A. S. Mayer, A. Klenner, A. R. Johnson, K. Luke, M. R. E. Lamont, Y. Okawachi, M. Lipson, A. L. Gaeta, and U. Keller, Opt. Express **23,** 15440-15451 (2015).

# Supplementary material for:
# Mid-infrared frequency comb with 25 pJ threshold via CW-seeded optical parametric generation in nonlinear waveguide

Mikhail Roiz, Jui-Yu Lai, Juho Karhu & Markku Vainio

**Supplementary note 1: improved experimental setup and relative CEO comparison**

In addition to applying a waveguide crystal instead of a bulk crystal for OPG, we have improved the experimental setup compared to our previous reports [1] by eliminating optical fibers in the CW seed laser phase-locking procedure. In Fig. S1 (a) one can see the old scheme that included two fiber couplers/splitters to phase-lock the CW seed laser to the pump comb via supercontinuum. The phase-locking procedure is needed to reference the CW seed laser to the pump comb, ensuring stabilization of the Carrier-Envelope Offset (CEO) frequency of the mid-infrared comb. The CW seed laser output was divided into two arms: one arm was used for OPG seeding, while the other arm was used to phase-lock the seed laser. These two arms had two different and long paths in optical fibers. The additional phase noise accumulated in the fibers due to acoustic noise and fiber temperature drifts [2] negatively affected the relative stability of the CW seed laser against the pump laser, inducing small fluctuations on the idler CEO over time. This effect is clearly visible when we compare the bulk and waveguide CW-seeded OPG idler comb offset frequencies using the old and new schemes (Fig. S2). The details of this comparison are explained below but, in brief, the frequency of beat note between the two idler combs was measured with a frequency counter over several minutes. In the case of old scheme, the frequency counting experiment and the corresponding Allan deviation plot (Fig. S2 (a)) reveal excess fluctuations between the idler CEOs of the bulk and waveguide OPG combs. Fiber noise was confirmed to be the main origin of these fluctuations; the fluctuations were greatly amplified when the fibers were touched.

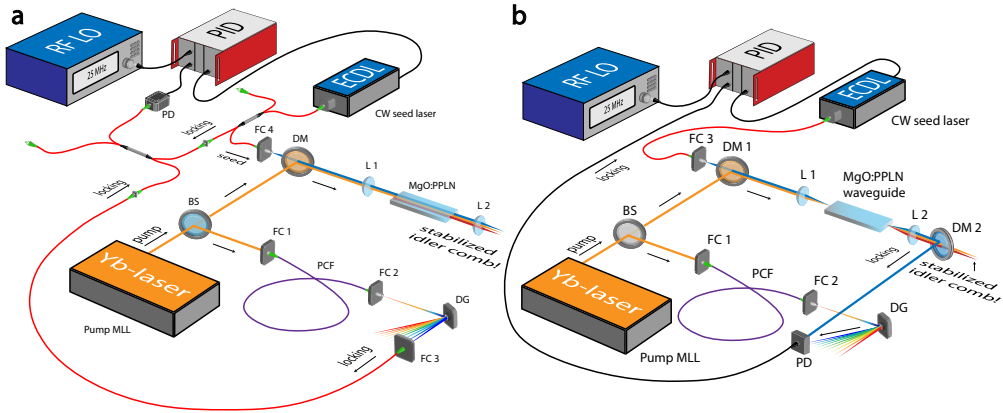

Fig. S1. Experimental setups of the old (a) and new (b) phase-locking schemes for the CW-seeded OPG; The main difference between the old and new schemes is elimination of optical fibers, which helps to remove drifts of the idler CEO; BS: beam splitter, FC: fiber collimator, DG: diffraction grating, DM: dichroic mirror, L: lens, PD: photodiode, PID: Proportional–Integral–Derivative controller, ECDL: External Cavity Diode Laser, PCF: Photonic Crystal Fiber.

In order to solve the fiber-noise issue, in the present work, we implemented phase-locking using the CW seed laser light extracted immediately after the OPG crystal output (Fig. S1 (b)). Effectively, this scheme stabilizes the CW seed laser phase to the pump comb *at the OPG* *crystal* significant noise reduction in the comparison of idler CEOs of the bulk and waveguide OPG combs. Indeed, this is another advantage of our scheme, since the phase-locking point (determined by the position of the photodetector in the phase-locking setup) can be chosen freely.

In Fig. S3. one can see the experimental setup that was used to compare the relative CEO stability between the bulk and waveguide CW-seeded OPGs. The *amplified* pump laser (up to 10 W of average power at maximum) was used to drive both the bulk and waveguide CW-seeded OPGs, however we used two independent CW seed lasers for seeding. Referring to Fig. S1, both CW lasers are phase-locked *after the nonlinear crystal/waveguide*, which allows the phase-locking to compensate for extra noise and maintain a better relative stability between the two CW seed lasers (at the same time keeping the absolute stability as well). We used two different local oscillators (RF LO 1 and RF LO 2) for phase-locking of two independent seed CW lasers. Since the RF LO defines the idler CEO, their frequency difference was set to 30 MHz leading to 30 MHz target frequency for the frequency counting experiment shown in Fig. S2. A simple delay line is used to match the idler pulses in time for their relative CEO stability measurement.

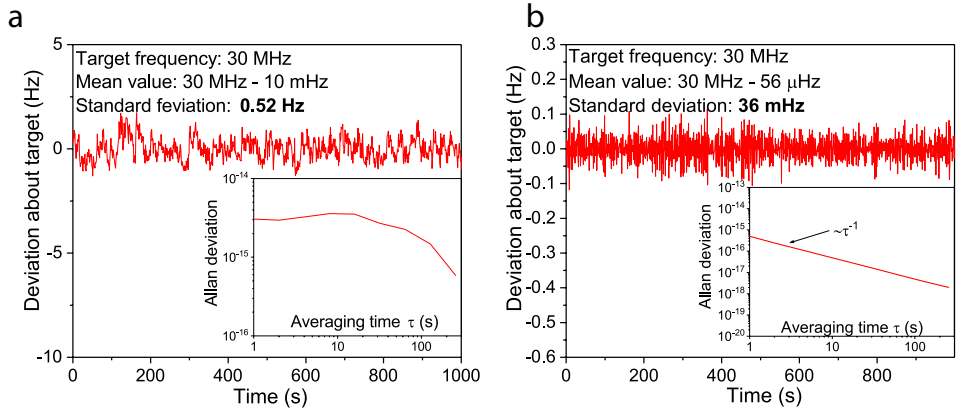

Fig. S2. (a) Frequency counting experiment of the relative offset frequency beat note between the waveguide and bulk CW-seeded OPG idler combs in the case of the old scheme; (b) same as (a) but with the new scheme; counter gate time is 1 second for both measurements.

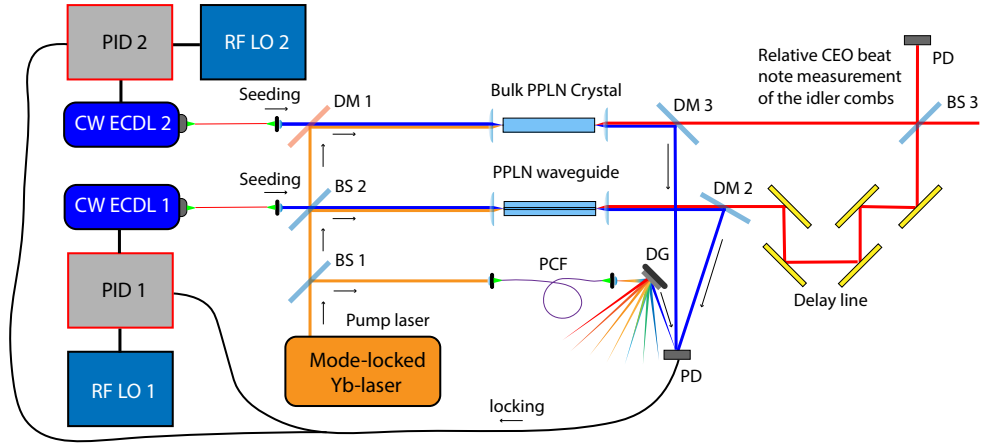

Fig. S3. Experimental setup for relative CEO comparison between the idler combs produced in the bulk and waveguide CW-seeded OPGs; BS: beam splitter, FC: fiber collimator, DG: diffraction grating, DM: dichroic mirror, L: lens, PD: photodiode, PID: Proportional–Integral–Derivative controller, ECDL: External Cavity Diode Laser, PCF: Photonic Crystal Fiber.

We also measured PSD phase noise of the beat note shown in Fig. 4c of the main text. The result can be seen in Fig. S4, which demonstrates very low integrated phase noise of just 22.6 mrad. It is worth noting though that this phase noise measurement does not exactly correspond to the actual offset frequency phase noise of the MIR comb, since this measurement does not include phase fluctuations of the SC. In order to see the SC phase fluctuations this measurement has to be done by heterodyning the MIR comb with a stable CW reference laser (also in MIR), which we do not have available in our lab. Alternatively, one could measure the offset frequency phase noise using f-2f interferometry, but unfortunately our MIR comb does not span one octave, hence this method is also not applicable in our case.

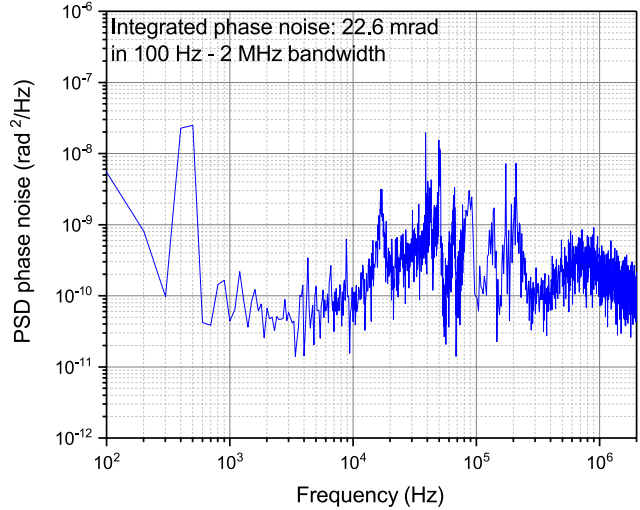

Fig. S4 PSD phase noise of the relative CEO beat note shown in Fig. 4c (main text)

**Supplementary note 2: pump broadening**

For our pump broadening experiment, we used a 3.7 cm long piece of photonic crystal fiber (NKT Photonics, NL-PM-750) that produced a supercontinuum near 1 µm wavelength with a typical spectrum demonstrated in Fig. S5. The regular pump spectrum without broadening is also depicted for comparison. Such broadening requires only about 70 mW of average input pump power inside the photonics crystal fiber for our laser system described in the main text.

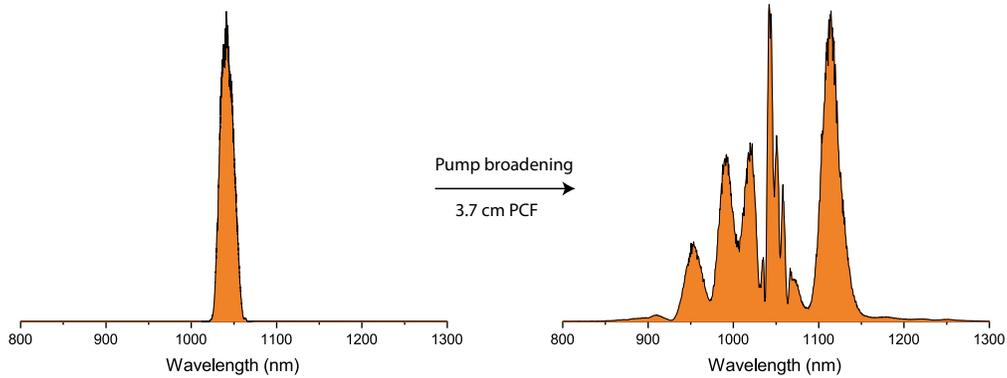

Fig. S5. Pump optical spectrum without broadening (left) and broadened in 3.7 cm photonic crystal fiber (right).

## References

1. M. Roiz, K. Kumar, J. Karhu, and M. Vainio, APL Photonics **6,** 026103 (2021).

2. L. Ma, P. Jungner, J. Ye, and J. L. Hall, Opt. Lett. **19,** 1777-1779 (1994).